\documentclass[journal]{IEEEtran}

\usepackage{lineno,hyperref}
\usepackage{units}
\usepackage{ifthen}
\usepackage{amsmath}
\usepackage{tikz}
\usepackage{booktabs}
\usepackage{graphicx}
\usepackage{tabularx}
\usepackage{multirow}
\usepackage{amsmath}
\usepackage{amsfonts}
\usepackage{amssymb}
\usepackage{arydshln}
\usepackage{setspace}
\usepackage{soul}

\usepackage{framed} % Framing content

\usepackage{multicol} % Multiple columns environment
\usepackage{nomencl} % Nomenclature package
\usetikzlibrary{shapes,snakes}
\usetikzlibrary{automata, positioning, arrows}
\ifCLASSINFOpdf
\else
\fi

\begin{document}
\title{A Data-Assisted Reliability Model for Carrier-Assisted Cold Data Storage Systems}

\author{Suayb S. Arslan,
        James Peng,
        and~Turguy Goker% <-this % stops a space
\thanks{Suayb S. Arslan is with the Department
of Computer Engineering, MEF University, Istanbul,
Turkey. e-mail: arslans@mef.edu.tr.}% <-this % stops a space
\thanks{James Peng ad Turguy Goker are with Advanced Channel Team, Quantum Corp., Irvine, CA, USA. e-mails: \{james.peng,turguy.goker\}@quantum.com.}}% <-this % stops a space

\markboth{UNEDITTED: Accepted for Publication in Elsevier Reliability and System Safety Journal, oct~2019}%
{Shell \MakeLowercase{\textit{et al.}}: Bare Demo of IEEEtran.cls for IEEE Journals}

\maketitle

\begin{abstract}
Cold data storage systems are  used to allow long term digital preservation for institutions' archive.  The common functionality among cold and warm/hot data storage is that the data is stored on some physical medium for read-back at a later time. However in cold storage, write and read operations are not necessarily done in the same exact geographical location. Hence, a third party assistance is typically utilized to bring together the medium and the drive. On the other hand, the reliability modeling of such a decomposed system poses few challenges that do not necessarily exist in other warm/hot storage alternatives such as fault detection and absence of the carrier, all totaling up to the data unavailability issues. In this paper, we propose a generalized non-homogenous Markov model that encompasses the aging of the carriers in order to address the requirements of today's cold data storage systems in which the data is encoded and spread across multiple nodes for the long-term data retention. We have derived useful lower/upper bounds on the overall system availability. Furthermore, the collected field data is used to estimate parameters of a Weibull distribution to accurately predict the lifetime of the carriers in an example scale-out setting. In this study, we numerically demonstrate the significance of carriers' presence and the key role that their timely maintenance plays on the long-term reliability and availability of the stored content.
\end{abstract}


\section*{Nomenclature}
\addcontentsline{toc}{section}{Nomenclature}
\begin{IEEEdescription}[\IEEEusemathlabelsep\IEEEsetlabelwidth{$V_1,V_2,V_3$}]
\item[$n$/$\Tilde{n}$]{The number of nodes present in the system / the blocklength of an MDS code.}
\item[$k$/$\Tilde{k}$]{The number of data nodes in the system / the payload length of an MDS code.}
\item[$s$]{Number of node states.}
\item[$\lambda$]{Rate of node failure process.}
\item[$\phi$]{Rate of robot (carrier) repair process.}
\item[$\mu$]{Rate of data repair process.}
\item[$\theta$]{Rate of failure detection process.}
\item[$N_s$]{Total number of Markov states when the number of node states is $s$.}
\item[ADF]{Availability,Detection,Failure.}
\item[RAID]{Redundant Array of Inexpensive/Independent Disks.}
\item[MTTDU]{Mean time to data unavailability.}
\item[MTTDL]{Mean time to data loss.}
\item[MDS]{Maximum Distance Separable.}
\item[UCER]{Uncorrectable error rate.}
\item[MTBF]{Mean time between failures.}
\item[CT]{Continous Time.}
\item[SBF]{Swaps/exchnges before failure.}
\item[TBS]{Time between swaps/exchanges.}
\item[TPM]{Transition Probability Matrix.}
\item[$\kappa$]{Tape cartridge damage ratio.}
\item[$\textbf{M}$]{Fundamental matrix of an absorbing Markov chain.}
\item[$\textbf{I}$]{Identity matrix.}
\item[$a_{ij}$]{The entry in the $i$th row and $j$th column of the matrix $\textbf{A}$.}
\item[$\epsilon$]{Drive read hard error probability.}
\item[$\Gamma(.)$]{Complete Gamma function.}
\item[$r$]{The rate of the MDS code given by $\Tilde{k}/\Tilde{n}$.}
\item[$\eta$]{Drive read hard error rate.}
\item[$B(.;.,.)$]{Incomplete Beta function.}
\item[CDF]{Cumulative Distribution Function.}
\item[$F_{\mathcal{P}}(.)$]{CDF of Poisson binomial distribution.}
\item[$\Re\{.\}$]{The real part of the argument.}
\item[$g$]{The shape parameter of the Weibull distribution.}
\item[$y$]{The scale parameter of the Weibull distribution.}
\item[$\beta_i$]{Survival probability of the node $i$.}
\item[$hs(.)$]{Harmonic sum function.}
\item[$LB$]{Performance Lower Bound.}
\item[$UB$]{Performance Upper Bound.}
\end{IEEEdescription}

% \fi
\IEEEpeerreviewmaketitle

\section{Introduction}

\IEEEPARstart{T}{he} physical medium on which the infrequently accessed data a.k.a. cold data is stored is quite susceptible to loss and corruption due to exposure to heat, humidity, dust-like contaminants, and in particular faulty and unpredictable read/write hardware behaviors. Typical examples include drives and tape/optical media where the data is laid on a physical medium for long term storage by utilizing the magnetic field directions \cite{allen2017materiality}. On the other hand, one of the most popular cold storage options in today's world is the Deoxyribonucleic acid (DNA) based media in which the data is embedded inside the DNA strands and read/write operations are  performed by special hardware doing specific chemical operations known as \textit{synthesizers} and \textit{sequencers}, respectively \cite{church2012next}. Strands are carried inside the bottles and protected in special environments such as the most recent \textit{molecular hopper} technology \cite{Qing2018} for long time preservation. Similarly,  data carriers (such as mechanical robots) are used to carry the tape cartridges or optic discs to drives in order to successfully complete read/write operations. 

In an accurate reliability modeling for cold data storage systems (such as tape), the media failure, faulty drive behaviour, failure detection, data spreading -- that is how much the data is spread across different independent cold storage nodes -- should all be taken into account \cite{arslan2014mds}. Particularly, the detection of faulty system behaviour and  lost data can be done by periodic system/data check operations which is known as \textit{scrubbing} in disk-based systems \cite{oprea2010clean}. Such a scrubbing process can be incorporated into reliability models \cite{ryu2009effects,schwarz2004disk,iliadis2008disk}. In contrast, in case of carrier failures, data is not lost but would be temporarily unavailable until the carrier is replaced/fixed by the maintenance team. Therefore, the actual reliability model should treat data durability and availability individually and be able to neatly merge these important data-related performance indicators. 

The reliability studies for disk arrays have been the center of research interest  in our modern age to help build sustainable and efficient data centers. One of the most comprehensive/generalization studies to date is given in \cite{Hafner2006} where hard errors and generic erasure coding are taken into account. Incorporation of both uncorrelated and correlated hard errors with Markov modelling are considered in a later study \cite{dholakia2008new}. Furthermore, some special cases of the proposed model in \cite{Hafner2006} is considered and analyzed in \cite{Paris2012}. One of the most general Markov model with horizontal and vertical data allocation policies is recently studied for disk systems \cite{Arslan2018} in which failure detection process is not taken into account. Furthermore, the analysis is extended to study the effect of multidimensional redundancy on aging \cite{arslan2014redundancy}. A special Markov chain is also considered to measure RAID-6 systems  \cite{rahman2018analysis} without much generalization. Due to analytical complexity, software tools are later introduced for accurate prediction of large scale data storage architectures \cite{hall2016tools}.  With regard to cold storage systems, several past research studied application-specific back-up systems in which keeping the copies of the data (replication) was the primary means of providing durability \cite{constantopoulos2005}. Although they have paid attention to failure detection problem, they did not configure their model for true cold storage environment requirements i.e., the necessity of carrying media from one location to another, detrimental effects of mechanical components, unavailability of carriers and drive-related hard errors. In \cite{HanChan}, the work is extended to cover 4-copy case where the backup system consists of both tapes and hard drives with different failure and repair rates i.e., a heterogeneous storage network. Since the storage media and internal mechanics are different for tape systems and hard drives, the proposed model quickly gets complicated as the number of copies increase. Therefore, extending it beyond 4-copy seems to be quite challenging and no systematic extension is proposed in the same line of work. In  \cite{ivanichkina2015}, a two dimensional Markov process is proposed for modeling explicit and latent errors in disk-based distributed storage systems in which failure detection is assumed to take almost no time. Since only disk-based systems are considered, the study is not extended to take into account the presence of carriers. 

Note that the overall scale-out system reliability relies on the durability of constituent cold storage units such as tapes. Several studies conducted life expectancy and media stability tests for magnetic tape and the results revealed that a theoretic 50-100 years lifespan is possible for magnetic tapes \cite{Judge2003,weiss2002}. Similar studies exist for optical disks \cite{shahani2003} as well. 

On the other hand, continuously failing storage media, conventionally neglected unavailability, unexpected and time-dependent hard errors (mostly due to aging) make these predicted numbers obsolete. In addition, in all kinds of digital data storage, data is usually erasure coded and  spread across different storage nodes through intelligent allocation schemes \cite{jernigan2009data} for better durability, space overhead and data loss characteristics \cite{cidon2013copysets}. Data access patterns and system-level details are collected by most companies for better understanding of their real workload and help develop better business strategies \cite{botezatu2016predicting}. This type of data can further be used using machine learning techniques to improve the overall storage reliability \cite{mahdisoltani2017proactive}. Hence, accurately estimating the data durability requires a data-assisted reliability model which should take all aforementioned detrimental effects into account.

Due to the complexities of data dependent density estimation and the time dependent carrier aging phenomenon, we formulated the proposed Markov model as a simulation platform to estimate the distributions of time to data-loss and data-unavailability. Simple statistics such as mean time to data loss (MTTDL) and unavailability (MTTDU) will be derived from the simulation data and compared for various choices of system parameters as well as few theoretical results known for a limited parameter space. For a better analysis, under reasonably good assumptions,  lower and upper bounds on the system performance will be derived and presented together with the numerical results. 

The rest of the paper is structured as follows. In Section II, the general multi-dimensional Markov model  is introduced in which carrier availability is taken care of. Moreover, its complexity analysis is addressed based on the number of states. In Section III, the most common problems in cold storage i.e., hard read errors and carrier unavailabilities are precisely modeled, transition and probability matrices are defined, lower and upped bounds on the performance are derived. Since the lifetime of the carrier is dependent on the usage pattern, data-assistance is utilized in Section IV to model the aging phenomenon. Finally, we provide few numerical results in Section V to demonstrate the significance of carrier availability and finally Section VI concludes the paper. 

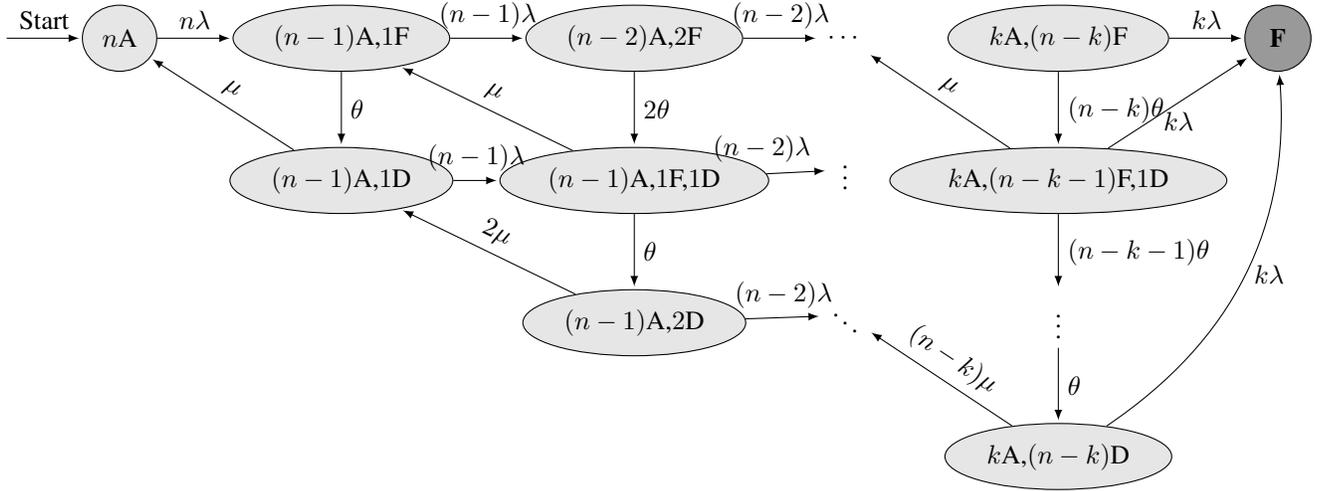
\begin{figure*}
\centering
%\hspace*{-2cm}
\begin{tikzpicture}
        % Setup the style for the states
        \tikzset{node style/.style={state, fill=gray!20!white}}
        \tikzset{node style2/.style={state, fill=gray!80!white}}
    
        \node[node style, ellipse]               (I)   {$n$A};
        \node[node style, ellipse, right=of I]   (II)  {$(n-1)$A,$1$F};
        \node[node style, ellipse, below=of II]  (IIb)  {$(n-1)$A,$1$D};
        \node[node style, ellipse, right=of II]  (III) {$(n-2)$A,$2$F};
        \node[node style, ellipse, below=of III]  (IIIb)  {$(n-1)$A,$1$F,$1$D};
        \node[node style, ellipse, below=of IIIb]  (IIIbb)  {$(n-1)$A,$2$D};
        \node[draw=none,  right=of III]   (IIIn) {$\cdots$};
        \node[draw=none,  ellipse, below=of IIIn]   (IIInb) {$\vdots$};
        \node[draw=none,  below=of IIInb]   (IIInbb) {$\ddots$};
        \node[node style, ellipse, right=of IIIn] (n)   {$k$A,$(n-k)$F};
        \node[node style, ellipse, below=of n]  (nb)  {$k$A,$(n-k-1)$F,$1$D};
        \node[draw=none, below=of nb]  (n-e) {$\vdots$}; 
        \node[node style, ellipse, below=of n-e] (neb) {$k$A,$(n-k)$D};
        \node[node style2, right=of n] (FF)   {\textbf{F}};
        %\node[draw=none, right=of n] (Failure)   {};
        
        % Draw empty nodes so we can connect them with arrows
        \node[draw=none, left=of I]   (a)    {};
    
    \draw[>=latex, auto=left, every loop]
         (a)   edge node {Start}                         (I)
         (II)  edge node {$\theta$}        (IIb)
         (IIb) edge node[sloped, above] {$\mu$} (I)
         (III) edge node {$2\theta$}       (IIIb)
         (IIb) edge node {$(n-1)\lambda$} (IIIb)
         (IIIb) edge node {$\theta$}       (IIIbb)
         (IIIb) edge node[sloped, above] {$\mu$} (II)
         (IIIb) edge node [very near end] {$(n-2)\lambda$} (IIInb)
         (IIIbb) edge node[sloped, above] {$2\mu$} (IIb)
         (IIIbb) edge node {$(n-2)\lambda$} (IIInbb)
         (n)   edge node {$(n-k)\theta$}                    (nb)
         (nb) edge node[sloped, above] {$\mu$}                    (IIIn)
         (nb) edge node[below] {$k\lambda$}                    (FF)
         (nb)   edge node {$(n-k-1)\theta$}                    (n-e)
         (n-e)   edge node {$\theta$}                    (neb)
         (neb) edge node[sloped, above] {$(n-k)\mu$}                    (IIInbb)
         (neb) edge[above, bend right, right=0.1] node {$k\lambda$} (FF)
         %(nb) edge node {$\lambda$}                (FF); 
         (I)   edge node {$n\lambda$}   (II)
         (II)  edge node {$(n-1)\lambda$}  (III)
         (III) edge node {$(n-2)\lambda$} (IIIn)
         (n) edge node {$k\lambda$}                (FF); 
         %
         %(n) edge node {$3t5t5t$}               (Failure);
         
    \end{tikzpicture}
    \caption{Generalized Markov model for $s = 3$ node states i.e., represented by Availability (A), Detection (D) and Failure (F). The state $\textbf{F}$ denotes the total failure/unavailability. Note that since $s=3$, corresponding Markov model can be represented on a two dimensional plane. In general, Markov model is $s-1$ dimensional.}
    \label{GMMs3}
\end{figure*}

\section{A Carrier-present Reliability Model}

In this section, we describe the general model that incorporates data as well as carrier presence in to a Markov model. We also consider a specific special case that is most common in realistic cold storage applications.

\subsection{General Model Description}

The proposed reliability model incorporates data--driven density estimation (a known distribution fit) and a continuous time Markov process to accurately estimate the density of the system data loss, and associated first order statistics such as MTTDL and MTTDU metrics. The model has two types of states: \textit{node} states and \textit{system} states all combined and characterized by the Markov states. Although the proposed model is generally applicable to any type of cold storage, we specifically consider a scale--out tape library system.

Although the paper shall treat the number of node states to be any arbitrary number, for simplicity, we will give our examples by proposing three different  node states that are the most commonly assumed in cold data storage community. There are defined to be \textit{Available} (A), \textit{Failed} (F) and \textit{Detected} (D) for a given node and we automatically generate the associated Markov states of the overall system. We would like to remind that typical continuous Markov processes are heavily used  for warm/hot storage devices  consisting of hard disk or solid state device arrays (such as given in \cite{Hafner2006}). Previous work focused on Markov processes with typically two node states, namely Available and Failure. Indeed, our treatment makes our model a generalized version of all the previous Markov models used with configurable parameters.  Although it becomes impossible to derive closed form expressions, we shall use approximations to derive analytical results for performance metrics such as lower and upper bounds on the mean statistics.

As assumed by many past research  studies such as \cite{burkhard1993}, \cite{wideman2016data} and \cite{suayb2019joint}, the cold storage system is protected by a $(\Tilde{n},\Tilde{k})$ Maximum Distance Separable (MDS) erasure code where $\Tilde{n}$ is the codeword (block) and $\Tilde{k}$ is the payload lengths, respectively. Note that with this setting, the conventional replication (copy) system corresponds to $\Tilde{k}=1$. The quantity $r = \Tilde{k}/\Tilde{n} = k/n$ is termed as the rate of the code. In other words, since we use MDS codes, we assume $\Tilde{n}$ and $\Tilde{k}$ to be a multiple of $n$ and $k$, respectively. Due to encode/decode complexity and without loss of generality, we shall assume $n = \Tilde{n}$ and $k = \Tilde{k}$ throughout the document to convey the main idea.

We realize that node states and Markov states are not the same, in fact, a Markov state can have triple node states. We can visualize each Markov state consisting of three buckets counting the number of nodes having each of A, F and D node states. Furthermore, we assume nodes are exactly the same type and fail with the same rate $\lambda$ i.e., homogeneous storage network. This means that due to a physical and irreversible error, data cannot be read from the tape that resides in that node. In addition, we have three more processes running in the system; two of them are the concurrent and identical data and carrier repairer processes and the third one is the concurrent and identical failure detector process. In this study, we assume that  an error detection process is run on tapes and repair them whenever an error is detected. In addition, media carriers such as robots can also be repaired since without their availability, all data operations will cease. Robot and data repairer as well as failure detector processes are assumed to be exponentially distributed with rates $\phi$, $\mu$ and $\theta$, respectively.

The complexity of the proposed continuous Markov process is strongly tied to the total number of system states which are a function of node states. For $s > 1$, i.e., number of node states being greater than one\footnote{In our formulation, we assume that node states A and F are naturally present in any reliability  model. In further extensions of the model proposed in this study, different processes can be incorporated such as detection, participation and aging.}, suppose we have $i$ available nodes with $k \leq i \leq n$ (each containing a single data chunk - remember this is a requirement for perfect data reconstruction) then we shall have $s-1$ node states to share a total of $n-i$ data chunks. In this case, the total number of decompositions is given by 
\begin{eqnarray}
\binom{n-i+s-2}{s-2}
\end{eqnarray}

For instance, Fig. \ref{GMMs3} shows all of the system states (Markov states) as a function of $k$ and $n$ if the node states are $s=3$, represented by A, D and F. Yet in another case, the state ``Queued for Service: QS" can be added to make number of node states $4$, i.e., $s=4$. The closed form expression to calculate the total number of Markov states (for general $s$) is given by (including the total failure state)
\begin{align}
N_s &= \sum_{i=k}^n \binom{n-i+s-2}{s-2} + 1 = \sum_{i=0}^{n-k} \binom{i+s-2}{s-2} + 1  \nonumber \\ &= \binom{n-k+s-1}{n-k} + 1 \label{nseqn}
\end{align}

Note that for the general case, we can further express $N_s$ as follows
\begin{eqnarray}
N_s &=& 1 + \binom{n-k+s-1}{n-k}  \label{eqn12}\\
&=& 1 + \binom{n-k+1}{1} + \sum_{i=2}^{s-1} \binom{n-k+i-1}{i} \label{eqn13}\\
&\geq& \sum_{i=0}^{s-1} \binom{n-k+1}{i}  \label{eqn14}
\end{eqnarray}
where we clearly see that equality in equation \eqref{eqn14} holds only for $s=2,3$. For small $s$,  equation \eqref{eqn14} can be used as an accurate approximation. Note that going from  \eqref{eqn12} to \eqref{eqn13}, we have used induction. In the \ref{appendixA}, we provide  tighter upper and lower bounds for $N_s$ and asymptotically analyze the complexity of the final Markov reliability model. 

\subsection{A special case: 3-node-state Markov Model}

\begin{table}[b!]
\normalsize
\centering
\begin{tabular}{|l|l|l|}
\hline \hline
Curr. State & Destinations     & Transition Rate \\ \hline
$i$A,$j$F,$z$D    & $(i-1)$A,$(j+1)$F,$z$D & $i\lambda$               \\ \hline
$i$A,$j$F,$z$D     & $i$A,$(j-1)$F,$(z+1)$D & $j\theta$               \\ \hline
$i$A,$j$F,$z$D     & $(i+1)$A,$j$F,$(z-1)$D & $z\mu$               \\ \hline \hline
\end{tabular}
\caption{State transitions and rates}
\label{statestable}
\end{table}

As introduced earlier, let us assume we have three node states: A, D and F. Therefore, there are only three (in general $s$) destinations that a next state change could result  in. For instance, for a given state index $(i$A,$j$D,$z$F$)$, Table \ref{statestable} summarizes all the state indexes as possible destinations. In the table, DS stands for destination state and $i,j,z$ should satisfy the inequalities
\begin{eqnarray}
k \leq i \leq n,  \ 
0 \leq j \leq n - k, \
0 \leq z \leq n - k,  \ 
i + j + z = n \nonumber
\end{eqnarray}

One of the things that is not accounted for in the simulation model is a transformation method from the three-index state name to a single index name that runs between 0 and $N_3-1$ for simulation convenience. One straightforward method is to let the common index $c_{ind}$ to be 
\begin{eqnarray}
c_{ind} &=& \frac{1}{2}(j+z)(j+z+1) + z + 1 \nonumber\\
&=& \frac{(n-i+1)!}{2(n-i-1)!} + z + 1 
= \binom{n-i+1}{2} + z + 1
\label{eqn3}
\end{eqnarray}
where $j+z = n-i$ is replaced to find the first equality. Note that there is a one-to-one relationship between $c_{ind}$ and $(i,j,z)$ and  we can similarly find the inverse transform of the index $c_{ind}$. To find ($i,j,z$) from a given $c_{ind}$, we first need to find the maximum $i \in \{k, k+1, \dots, n\}$ such that $c_{ind} > i(i+1)/2$. From equation \eqref{eqn3}, we can find $z$ since we know $n$ and $i$. Finally, we use the fact that $i+j+z = n$ to determine $j$. 

Also the maximum of $c_{ind}$ shall be achieved with the state index $(k,0,n-k)$. Since $N_s = \max\{c_{ind}\} + 1$, the following can be shown to be true
\begin{eqnarray}
N_3  &=&  \binom{n-k+1}{2} + n-k + 2 \\
&=& \frac{1}{2}
\left((n-k+1)(n-k) + 2(n-k+2)\right) \\
&=& \frac{1}{2}(n-k+2)(n-k+1) + 1 \\
&=& \sum_{i=1}^{n-k+1}i + 1 = \sum_{i=0}^{n-k}(i + 1) + 1
\end{eqnarray}

Note that the final equality is the same as the result given by the Eqn. \eqref{nseqn} when $s = 3$. In order to derive performance expressions and apply the general model to a particular practical application, we shall assume $s=3$ for the rest of our discussions.

\begin{center}
\begin{figure*}[htp!]
 \[
 \hspace*{-1.8cm}
 \textbf{Q} =  \left[ \begin{array}{ccccccc}
-n\lambda   & n\lambda\Delta_n & 0 & 0 & \dots & 0 & n\lambda(1-\Delta_n) \\
0 & -(\theta + (n-1)\lambda) & \theta & (n-1)\lambda\Delta_{n-1} &\dots & 0 & (n-1)\lambda(1-\Delta_{n-1}) \\
\mu & 0 & -(\mu + (n-1)\lambda) & 0 & \dots & 0 & (n-1)\lambda(1-\Delta_{n-1}) \\
0 & 0  & 0 & 0 & \dots & 0 & (n-2)\lambda(1-\Delta_{n-2}) \\
\vdots & \vdots & \vdots &  & \ddots & \vdots & \vdots \\
0 & 0 & 0 & \dots &   & -(k\lambda+(n-k)\mu) & k\lambda \\
0 & 0 & 0 & \dots & 0 & 0 & 0 \end{array} \right]\]
\normalsize
%\caption{Transition Rate Matrix given for the Markov model shown in Fig. \ref{GMMs4}.}
%\label{rate_matrix}
\end{figure*}
\end{center}

\section{Failure Types and Carrier Unavailability in Cold Storage}

Hard error scenarios are well understood in warm/hot storage realms i.e., when the drive and the storage medium are tightly coupled. In addition, there is no separate carrier availability problem due to this coupling. However, modeling and incorporating the hard errors as well as the carrier availability all at the same time into a reliability model is much more challenging in a cold data storage context.

Let us consider tape library systems as an example use case. One of the fundamental challenge is that the robots (carrier devices) can make some given number of exchanges--swaps (round-trips) before failure. Since such a constraint depends on time and the frequency of use (load of the system), this would add non-homogeneity to the Markov model at hand. Also, there are two driving forces for aging in the same system: (1) The user data access pattern which is usually less dominant in a cold storage setting and (2) the internally generated access requests due to system/data repair operations which will lead to extra robot exchanges, drive load/unload cycles, tape positioning etc., to be able to meet the system reliability goals. Similar observations can be made for other popular cold storage alternatives. 

In this study, we constrained our set and assumed three dominant factors two of which directly affects the data durability while the other only changes the data unavailability. These factors, which need to be identified carefully, are explained in detail in the following subsections.

\subsection{Drive Read Failure} This type of hard error is also pretty prevalent in warm/hot storage where drives become unable to read the data due to an uncorrectable error by the virtue of internal error correction decoding failure of the drive. Uncorrectable error rate (UCER) is usually given in terms of errors per number of bytes or bits read and are usually due to random noise effects. This hard error mechanism is usually assumed to be time-independent and seeing at least one read error can be calculated by independence assumption, given by
\begin{eqnarray}
\epsilon = 1 - (1 - UCER)^{tape\_capacity}
\end{eqnarray}
where we assume the worst case scenario i.e., bulk reads i.e., the entire tape is read and $tape\_capacity$ is the number of bytes a tape can store. Note that in cold storage, this worst case scenario is quite common.

\subsection{Storage Medium (Tape) Damage} In our study, we model the probability of tape damage due to manufacturing reasons (in the infant mortality period) at the onset or external factors such as humidity, pressure and stringent temperature conditions later on in their lifetime. Such factors are assumed to be static and persists after failure detection and correction throughout the lifespan of the data stored in the cold storage. For this rationale, we used $\kappa > 0$ to model this damage probability. More specifically, we refer to tape damages to be $\kappa \times 100$ percent of all the tapes contained in a given library.

\subsection{Carrier (Robot) Failure}

Suppose that library robots (carriers) can make $m$ number of exchanges (round-trips) before they fail and eventually become unable to complete tasks initiated by the libraries including the detection and repair processes. We assume the robot types and qualities are identical in all of the libraries. Based on the available data extracted from our local library systems, we shall show that $m$ can be modelled as a  Weibull distributed random variable with some shape($g$) and scale($y$) parameters.

It is typical to assume the time between exchanges to be exponentially distributed  with rate $\omega$. The rate $\omega$ depends on a number of parameters such as the number of users using the system, the total number of libraries in a scale-out setting, the time of the year etc. The time to robot failure (characterized by the random variable $Y$) is therefore the sum of $m$ exponential distributions each with the same rate $\omega$ i.e., Gamma distributed with the following pdf
\begin{align}
f_{Y}(y; \omega, m) = \frac{\omega^m y^{m-1}e^{-\omega y}}{\Gamma(m)} = \frac{\omega^m y^{m-1}e^{-\omega y}}{(m-1)!} \ \ \omega,m > 0.
\end{align}
where $\Gamma(.)$ is the complete Gamma function. Note that since Gamma distribution is NOT memoryless, it requires aging to be taken care of by inserting the time dependent conditional CDF (carrier survival probabilities) given by
\begin{eqnarray}
\beta(t) &=& P(Y > t | l \textrm{ exchanges made}) \nonumber \\ &=& 1 - P(Y < t | l \textrm{ exchanges made}) \nonumber \\ 
&=& \frac{\Gamma(l, \omega t)}{\Gamma(l)} = \frac{\Gamma(l, \omega t)}{(l-1)!}, \ \ \ t > 0.
\end{eqnarray}
where $l < m$ exchanges are assumed to be made by the robot. In that case, the latter conditional probability is also Gamma distributed with the pdf $f_Y(y; \omega, l)$. Also, the included upper incomplete Gamma function is given by
\begin{eqnarray}
\Gamma(l, \omega t) = \int_{\omega t}^{\infty} x^{l-1} e^{-x} dx 
\end{eqnarray}

On the other hand, the hard error rate is given by $\eta = 1 - (1-\epsilon)(1-\kappa)$. Note that we need to modify the model to compensate for the hard errors.  Note that hard errors split the state transition from the current state $(i$A,$j$F,$z$D$)$ to the destination state $((i-1)$A,$(j+1)$F,$z$D$)$ which originally happens with rate $i\lambda$. Similar to the observations in previous studies, for each state with $i > k$, we need a transition to the total failure ($\textbf{F}$) state to be able to incorporate the hard errors.  With $i$ available nodes, we can tolerate up to $i - k - 1$ concurrent hard errors to successfully make it to the state $((i-1)$A,$(j+1)$F,$z$D$)$. Assuming independence, this happens with probability
\begin{eqnarray}
\Delta_i := \sum_{l=0}^{i-k-1}  \binom{i}{l}\eta^l(1-\eta)^{i-l} 
= 1 - I_{\eta}(i-k, k+1)
\end{eqnarray}
where $l$ is the number of hard errors that occur at the same time while rebuilding or during regular data checks. Also we have used the regularized beta function $I_x(a,b) = B(x; a, b)/B(a,b)$ instead to avoid the instability and precision issues of the binomial CDF. Here $B(x; a,b)$ is called the incomplete beta function and is given by
\begin{eqnarray}
B(x; a,b) = \int_{0}^x v^{a-1} (1-v)^{b-1}dv
\end{eqnarray}
and its complete version $B(a,b) = B(1; a,b)= \frac{\Gamma(a)\Gamma(b)}{\Gamma(a+b)}$. Note that we have the following limit
$\lim_{\eta \rightarrow 1} \Delta_i = 0 $
where the proposed model reduces to a simple transition from ($n$A,$0$F,$0$D) to $\textbf{F}$ with rate $n \lambda$. Since the hold times are assumed to be exponential, the mean time to stay in that state is $1/n\lambda$ which can be thought as the lower bound on the durability of the system. On the other hand, for small $\kappa$ and $\epsilon$, the upper bound can closely be approximated by the reliability of the system presented in Fig. \ref{GMMs3}. Finally, we summarize the new state transition table in Table \ref{dest_table2} with indexes satisfying
\begin{eqnarray}
k + 1 \leq i \leq n,  \ \ \
0 \leq j \leq n - k -1 \nonumber \\ 
0 \leq z \leq n - k - 1,  \ \ \
i + j + z = n. \nonumber
\end{eqnarray}

Note that if we redraw the overall Markov system given in Fig. \ref{GMMs3} to incorporate hard errors,  it will make it look more complicated. To reach such a transition table, we have made a few assumptions that can be listed as follows.
\begin{itemize}
    \item In a typical repair process, only $k$ tapes are selected for repair process. In case of locally repairable codes \cite{LRC}, this number can be reduced. Alternatively, more than $k$ tapes can be requested and only earliest $k$ reads can be used to improve performance. 
    \item Hard errors are assumed to be independent i.e., data cannot have more than one segment(data chunk)  in the same library node.
\end{itemize}

\begin{table}[t!]
\normalsize
\centering
\begin{tabular}{|l|l|l|}
\hline \hline
Curr. State & Destinations     & Transition Rate \\ \hline
$i$A,$j$F,$z$D    & $(i-1)$A,$(j+1)$F,$z$D & $i\lambda \Delta_i$               \\ \hline
$i$A,$j$F,$z$D    & \textbf{F} & $i\lambda(1-\Delta_i)$             \\ \hline
$i$A,$j$F,$z$D     & $i$A,$(j-1)$F,$(z+1)$D & $j\theta$               \\ \hline
$i$A,$j$F,$z$D     & $(i+1)$A,$j$F,$(z-1)$D & $z\mu$               \\ \hline \hline
\end{tabular}
\caption{State transitions and corresponding transition rates}
\label{dest_table2}
\end{table}

\subsection{Transition Rate and Probability Matrices}

The transition rate matrix (TRM) $\textbf{Q}$ is a $N_s \times N_s$ real valued matrix, whose entries $q_{ij} \geq 0$ for $i \not= j$ represent the rate departing from state $i$ and arriving in state $j$. The transition rate matrix for our generalized model is shown below. Note that we have included the total failure state $(\textbf{F})$ as part of the matrix and hence the last row becomes all-zero vector. We notice that the diagonal entries satisfy
\begin{eqnarray}
q_{ii} = - \sum_{j \not= i} q_{ij} \Rightarrow  \sum_{j} q_{ij} = 0 \ \ \forall i\in\{0,1,\dots,N_s-1\}
\end{eqnarray}
which means the rows of the matrix must sum to zero. 

Furthermore, let us define $\overline{\textbf{Q}}$ in which entries are defined as $\overline{q}_{ij} := q_{ij}/|q_{ii}|$ for all $i$ and $j$. Then, the transition probability matrix (TPM) is given by $\textbf{P} = \textbf{I} + \overline{\textbf{Q}}$. Note that this is the precise version of \textit{uniformization} technique\footnote{In that uniformization technique, $q_{ii}$ is replaced with $\gamma \geq \max |q_{ii}|$.} that compute transient solutions of finite state continuous-time Markov chains, by approximating the process using a discrete time Markov chain. This formulation will be useful when we derive the upper bound on the performance. 

\subsection{Modeling the Carrier Availability}

The treatment of the previous section did not include the availability of the carriers in the state transition matrix. One of the observations is that although the node failures (e.g. tape failures) are independent of robots' availability, node repair and failure detection mechanisms are highly dependent on the availability of carriers (robots) i.e., the rates that describe failure detection and node repair must be time-dependent as well. As the time passes by, detection and repair rates will go down unless carriers are updated sufficiently fast. 

When nodes fail due to various reasons, failure detection process  immediately commences. Similarly, when these failures are detected, the associated repair process starts immediately. So for a given operating time $t$, system robots will not be of the same age and quality (due to potential replacements etc). This leads to unequal treatment of storage nodes and our simulation setup must keep track of indexes for which robots are replaced in order to  model the aging phenomenon. 

\subsubsection{Time-dependent Failure Detection}

For simplicity, let us assume each node has a single carrier (through averaging arguments, it can be generalized to multiple carriers without changing the following discussion) and let $\psi_o(t)$ be the probability of $o  \in \{0,1,\dots,i\}$ available robots (conditioned on a specific set of $i$ nodes) in the system at time $t$ with survival probabilities $\beta_{s_1},\beta_{s_2}, \dots, \beta_{s_i}$ where $s_m \in \{1,2,\dots,n\}$. It can be shown that 
\begin{eqnarray}
\psi_o(t) = \Re\{F_{\mathcal{P}}(o)\} - \Re\{F_{\mathcal{P}}(o)\}
\end{eqnarray}
where $\Re\{.\}$ denotes the real part and $F_{\mathcal{P}}(.)$ is the  CDF of Poisson binomial distribution given by
\begin{align}
F_{\mathcal{P}}(o) = \frac{1}{n+1} \sum_{l=0}^i e^{-\frac{2\sqrt{-1}\pi l o}{n+1}} \prod_{m=1}^{i} (1 + (e^{\frac{2\sqrt{-1}\pi l}{n+1}}-1)\beta_{s_m}(t))
\end{align}
where $\sqrt{-1}$ is the complex number that is a solution of the equation $x^2 = -1$. On the other hand, since in our study we assume failures, detections and repairs all to be exponentially distributed and detection and carrier repairs can only happen consecutively, the natural consequence of sum of multiple independent exponential distributions is no surprise. However, to be able to make our later analysis analytically tractable, we will use a first order approximation in this subsection\footnote{Although in numerical result section, we will show that this assumption is a good approximation by simulating the actual distributions.}. More specifically, we will assume the sum of $x+1$ exponential distributions with rates $\phi,\theta_1,\dots,\theta_x$ to be approximately exponentially distributed with rate $R_{\boldsymbol{\theta}}(\phi)$ given by
\begin{eqnarray}
R_{\boldsymbol{\theta}}(\phi) = \frac{1}{\frac{1}{\phi} + \sum_{c=1}^x \frac{1}{\theta_c}} \label{exptailapprox1}
\end{eqnarray}
where $\boldsymbol{\theta} = [\theta_1, \dots, \theta_x]$. When a failure event is detected by the system, a state transition happens from the originator state $(i$A,$j$F,$z$D$)$ to the destination state $(i$A,$(j-1)$F,$(z+1)$D$)$ for $j>0$. While performing the detection, we need $j$ robots to complete the process, if found less, say $b < j$, then we need to repair $j-b$ robots to have a total of $j$ robots to work on the detection process. Suppose that at time $t$, we condition on having $l$ failed robots satisfying $0 \leq l \leq j \leq n - k$. Then the conditional repair rate i.e., the rate of making the detection transition in the Markov model is given by
$(j-l)\theta + lR_{\boldsymbol{\theta}}(\phi)$. Thus, summing over all possibilities of $l$, we get the unconditional node failure detection rate given by
\begin{eqnarray}
\theta_j(t;\phi) &=& \sum_{l=0}^j \left((j-l)\theta + lR_{\boldsymbol{\theta}}(\phi)\right) \psi_{j-l}(t) \\
&=& j\theta - (\theta - R_{\boldsymbol{\theta}}(\phi))\sum_{l=0}^j l\psi_{j-l}(t)  \\
&=& j\theta - (\theta - R_{\boldsymbol{\theta}}(\phi))\sum_{m=1}^j (1-\beta_{s_m}(t)) \\
&=& j\theta - \frac{\theta^2}{\theta + \phi} \sum_{m=1}^{j} (1-\beta_{s_m}(t))
\end{eqnarray}
where $s_m \in \{1,\dots,n\}$. Notice that we have the inequality for any $t$,
\begin{eqnarray}
j\theta - \frac{j\theta^2}{\theta+\phi} \leq \theta_j(t;\phi) \leq j\theta
\end{eqnarray}
which implies that as $\phi \rightarrow \infty$ i.e., robot repairs being instantaneous, the detection rate would be $j\theta$ which is the same as that of without any robot failures as given in Fig. \ref{GMMs3}. 

\subsubsection{Time-dependent Carrier Repair} After a node failure is detected, our system immediately begins the repair process and the completion of the repair process implies a state transition  from the originator state $(i$A,$j$F,$z$D$)$ to the destination state $((i+1)$A,$j$F,$(z-1)$D$)$ for all originator states having $z>0$.

Let us suppose we are in state $(i$A,$j$F,$z$D$)$ at time $t$ and $l$ of $i$ available nodes have their carrier robot already failed. Note that for classical MDS codes, we need to have $k$ helper nodes to be able to complete the data request successfully\footnote{Various network codes exist that may require to access more than or less than $k$ helper nodes with partial node content accesses for full recovery \cite{dimakis2011survey}. The present discussion only slightly changes in case such class of codes are used instead.}. Suppose further that $x$ of these requests are from the failed set, and $k-x$ are from the available and operational ones. Due to sampling without replacement, probability of that happening is given by the hypergeometric distribution\footnote{Sampling with replacement would lead to a Binomially distributed statistics instead.}. In this particular condition, we need to wait for the $x$ failed carriers to be repaired first which is given by the maximum repair time and typically not distributed exponentially. In fact, this distribution can be shown to be equal to the sum of exponential distributions which in this subsection is assumed to be close to another exponential distribution with rate $1/\phi \sum_{m=1}^x 1/m$ where the harmonic sum in the rate can be approximated closely by
\begin{eqnarray}
hs(x) := \sum_{m=1}^x \frac{1}{m} \approx \log(x) + \zeta + \frac{1}{2x} - \frac{1}{12x^2} + \frac{1}{120 x^4} \label{eq_approx}
\end{eqnarray}
where $\zeta = 0.5772156649$ is known as Euler--Mascheroni constant. 

After all the necessary repair information is collected by any of the $z$ detected nodes, each begins the computation needed for the repair process and write the repaired data to the corresponding storage unit. But the write process needs at least one carrier/robot available. The availability analysis is quite similar to the same case with detection process (each node uses their own robot for detecting the failure) and thus the rate of such happening is \st{given by} represented by $\mu_z(t;\phi)$ expressed as
\begin{eqnarray}
\mu_z(t;\phi) = z\mu - \frac{\mu^2}{\mu + \phi} \sum_{m=1}^{j} (1-\beta_{s_m}(t))
\end{eqnarray}

On the other hand, the conditional repair rate (conditioned on $x$ and $l$) can be expressed as
\begin{align}
\mu_{iz}(t;\phi,k | x, l) = \frac{\binom{i-l}{k-x}\binom{l}{x}}{\binom{i}{k}} \left( \frac{1}{\mu_z(t;\phi)} + \frac{hs(x)}{\phi}  \right)^{-1}
\end{align}
where $s_m \in \{1,\dots,n\}$. Finally, the unconditional repair rate can be obtained by summing over all $x$ and $l$ as follows,
\begin{align}
    \mu_{iz}(t;\phi,k) &= \sum_{l=0}^i \psi_{i-l}(t) \sum_{x=0}^l \mu_{iz}(t;\phi,k | x, l) \\
    &= \sum_{l=0}^i \psi_{i-l}(t) \mu_z(t;\phi) \\ & + \sum_{l=0}^i \psi_{i-l}(t) \sum_{x=1}^l\mu_{iz}(t:\phi,k | x, l) 
\end{align}

Note that if $\phi \rightarrow \infty$, i.e., we assume immediate robot repairs, we shall have 
\begin{align}
    \mu_{iz}(t;\infty,k) &= \sum_{l=0}^i \psi_{i-l}(t) \sum_{x=0}^l \frac{\binom{i-l}{k-x}\binom{l}{x}}{\binom{i}{k}} z\mu = z \mu
\end{align}
meaning that robot  repairs  being instantaneous,  the  node repair  rate  would  be $z\mu$ which  is  the same as that of without any robot failures. Finally, we summarize the new state transition table in Table \ref{dest_table3} with indexes satisfying the following inequalities
\begin{eqnarray}
k \leq i \leq n,  \ 
0 \leq j \leq n - k \ 
0 \leq z \leq n - k,  \ 
i + j + z = n \nonumber
\end{eqnarray}

\begin{figure*}[t!]
\centering
  \includegraphics[width=0.8\linewidth]{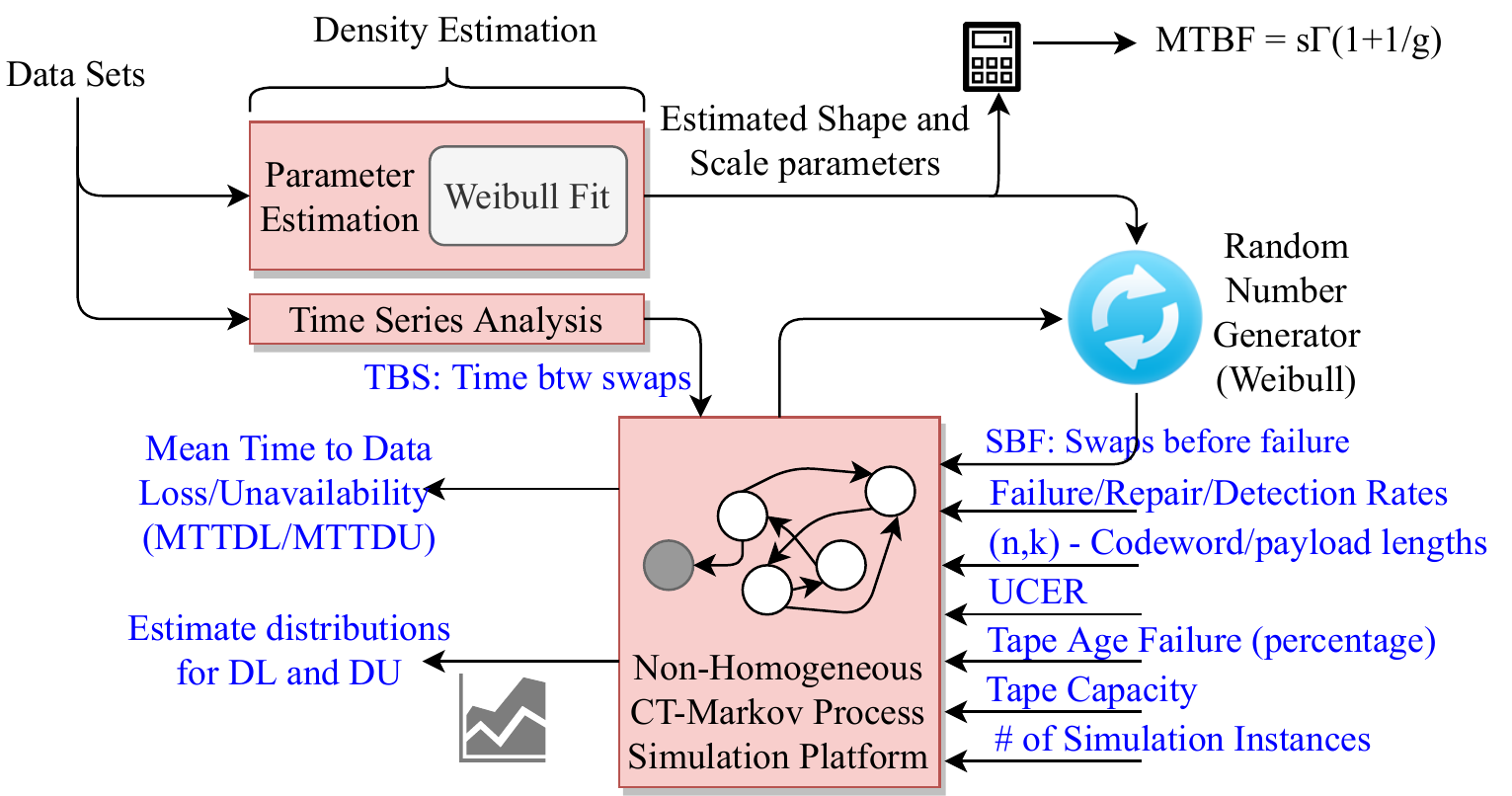}
  \caption{Data-Inspired overall modeling and simulation platform implemented in MATLAB. Density estimation problem is reduced down to parameter estimation through assuming a Weibull distribution for robot exchange performance data. In this figure, $s$ represents the scale and $g$ represents the shape parameters. MSBF: Mean Time Before Failure.}
  \label{fig:modelling}
\end{figure*}

\subsection{Lower/Upper bounds on the Performance}

For a given finite carrier repair rate $\phi < \infty $, if we let the exchange rate tend to large values the carrier repairs will not be able to catch up,  eventually resulting in total carrier unavailability. In that particular case, it is of interest to drive the lower bound on performance in a closed form. We note that in case of total carrier unavailability,  there is no failure detection and therefore it means no data repair in a cold storage context and hence, the survival time depends on which state the system is in and whether the hard error leads to unrecoverable state transitions. In light of this observation, the lower bound ($LB$)  is approximated in \ref{appendixB} in terms of $\Delta_i$'s as follows
\begin{eqnarray}
LB &\approx \sum_{i=k}^n (1 - \Delta_i) \prod_{j=i+1}^n \Delta_j  \Bigg( \log\left(\frac{n}{i-1}\right)^{1/\lambda}  \nonumber \\ &+ \frac{1}{\lambda}\sum_{l=0}^1(-1)^l\left(\frac{1}{2n_l} - \frac{1}{12n_l^2} + \frac{1}{120n_l^4}\right) \Bigg) \label{lowerbound}
\end{eqnarray}
with $\Delta_k = 0$ and $n_l = n - l(n-i+1)$. On the other hand, if we let the exchange rate tend to  zero there will be no need for  carrier  repairs,   resulting  in  total carrier availability. In that particular case, it is of interest to drive the  upper bound on performance  in  a  closed form. We realize that there is only one absorbing state in our model (Failure state) and hence, the TPM is already in its canonical form,
\[ \textbf{P}_{N_s \times N_s} = \left[
    \begin{array}{c;{3pt/3pt}r}
    \mbox{$\textbf{L}_{N_s-1 \times N_s-1}$} & \mbox{ $\textbf{R}_{N_s-1 \times 1}$} %\begin{matrix} 0 \\ 0 %\end{matrix} 
    \\ \hdashline[3pt/3pt]
    \begin{matrix} 0 & \dots & 0 \end{matrix} &  1
    \end{array}
    \right]
\]

For an absorbing Markov chain, we know that the inverse of $\textbf{I} - \textbf{L}$ matrix is called the fundamental matrix (denoted as $\textbf{M}$) and it can be expressed as
\begin{eqnarray}
\textbf{M} = (\textbf{I}-\textbf{L})^{-1} = \textbf{I} + \sum_{i=1}^{\infty} \textbf{L}^i
\end{eqnarray}
in which $m_{ij}$ entry provides the expected number of times that the Markov process visits the transient state $s_j$
when it is initialized in the transient state $s_i$. Since we initially assume all nodes to be available in the beginning, we are interested in $m_{1j}$s i.e., the system is assumed to be in state $n$A in the beginning of the operation. Since for $s_j$, all outgoing transitions happen according to exponential distributions and the hold time is given by the minimum which  is also distributed exponentially with rate $-q_{jj}$. This implies the average hold time in each visit to $s_j$ is given by $-1/q_{jj}$. Finally, the upper bound can be approximated by
\begin{eqnarray}
UB \approx -\sum_{j=1}^{N_s-1} \frac{m_{1j}}{q_{jj}} \label{upperbound}
\end{eqnarray}

Note that this is only an approximation since TPM is an approximation to the continuous time Markov model. Also, we can analytically assess the upper bound on the time-dependent performance including the robot failure and repair processes by considering only the rate matrix given by Table \ref{dest_table3} instead of Table \ref{dest_table2}. This is possible because we have approximated distributions as exponential to keep Markovianity intact. This approximation will later in numerical results section be verified to be sufficiently accurate for the range of parameters of interest.

\begin{table}[t!]
\normalsize
\centering
\begin{tabular}{|l|l|l|}
\hline \hline
Current State & Destination State     & Transition Rate \\ \hline
$i$A,$j$F,$z$D    & $(i-1)$A,$(j+1)$F,$z$D & $i\lambda \Delta_i$               \\ \hline
$i$A,$j$F,$z$D    & \textbf{F} & $i\lambda(1-\Delta_i)$             \\ \hline
$i$A,$j$F,$z$D     & $i$A,$(j-1)$F,$(z+1)$D & $\theta_j(t;\phi)$               \\ \hline
$i$A,$j$F,$z$D     & $(i+1)$A,$j$F,$(z-1)$D & $\mu_{iz}(t;\phi,k)$               \\ \hline \hline
\end{tabular}
\caption{State transitions and corresponding transition rates.}
\label{dest_table3}
\end{table}

\section{Data-Assisted Modeling Framework}

In our modeling framework, we utilize a data-inspired approach for estimating the number of round-trips (exchanges in our context) that a carrier make before a critical failure happens. The critical failure takes place when the robot is no longer able to operate within the library system due to various reasons till they are replaced with the new one. In our tape application, the total number of robot exchanges before failure (SBF) is assumed to be Weibull distributed which shall be validated by the collected field data using enterprise Quantum libraries. Weibull distribution is completely characterized by two independent parameters called the shape ($g$) and scale ($y$). The reason we choose Weibull is  twofold. First, it is the generalization of the most commonly assumed exponential distribution (single parameter) in literature. In other words, by selecting appropriate parameter values Weibull can be transformed to exponential distribution. Secondly, it is heavy tailed and closely characterize the observed field data. We realize that the heavy-tailed distributions characterize various types of data accurately as the number parameters of the distribution increase. For instance, it is reported in various studies  that the data object size tends to possess heavy-tailed distribution such as Pareto \cite{Satyanarayanan}. On the other hand, several studies show that heavy-tail distributions might well characterize local file system dynamics and file sizes \cite{downey2001structural}, archival data \cite{ramaswami2014modeling} and the data stored and communicated over the world wide web \cite{gong2001tails}. 

We note that based on the available field data and Weibull assumption, the challenging density estimation problem is transformed into parameter estimation problem. More precisely, we estimate the shape ($g$) and scale ($y$) parameters of the distribution through simple linear regression. Secondly, we obtain an estimate of the distribution of the time between exchanges/swaps. Using the same data set, this distribution is observed to have exponential tail and hence a single parameter (the rate) will have to be estimated. An exponential assumption is also quite nifty because the corresponding count process will become analytically tractable Poisson distribution. Since the estimated parameter is a function of the utilization rate of the system and hence is time-dependent, we shall test a range of values in our simulations to illustrate the overall picture. A summary of the modeling framework is depicted in Fig. \ref{fig:modelling}. In this framework, the data-based parameter estimations ($\hat{g}$ and $\hat{y}$) are fed into the proposed non-homogeneous Markov Process as estimated inputs.  In addition to these inputs, we also set the rest of the simulation parameters $\lambda,\mu,\theta,n,k,\kappa,\epsilon$ as well as the number of simulation instances to some appropriate values based on the field data and our experience with 6TB tapes. The system is protected with a $(n,k)$ MDS code. The random number generator chooses a random SBF value according to the estimated Weibull distribution and repeats this process and uses a unique realization at each iteration of the simulation. We typically simulate over 10000 times to obtain reliable values. 

The main purpose of the simulation platform is to estimate the distribution of the overall data loss and/or unavailability (which ever one  degrades the performance first) at the same time to demonstrate the implicit relationship of these two important performance metrics. In other words, we can finally numerically estimate MTTDL and data MTTDU metrics quite confidently. In addition, the mean value of the number of exchanges is given by $\hat{y}\Gamma(1 + 1/\hat{g})$ which shall be used as the guideline of robot performance in the numerical results section.

\begin{figure}[t!]
\centering
  \includegraphics[width=\linewidth]{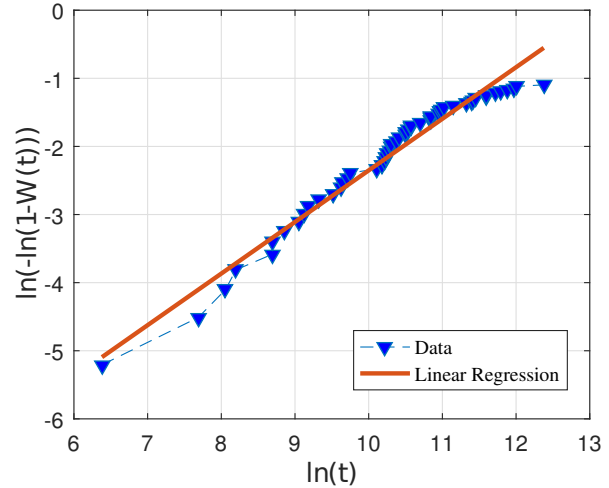}
  \caption{Robot exchange data and Weibull parameter estimations using a linear regression.}
  \label{fig:data_robot_xph}
\end{figure}

Note that there are more than one way for the estimation of the Weibull distribution parameters i.e., $g$ and $y$. We adapt simple linear regression in this study to estimate these parameters of the Weibull distribution. However, few algebraic manipulations are needed to put the CDF of the Weibull in an appropriate form. Accordingly, let us remember the Weibull CDF $W(t)$ as given by the equation.
\begin{eqnarray}
W(t) = 1 - e^{(t/y)^g}
\end{eqnarray}
which can be rearranged and expressed as the following linear equation
\begin{eqnarray} \ln\left(-\ln\left(1-W(t)\right)\right) =  g\ln(t/y) =  g\ln(t) - g\ln(y) \label{eqn411}
\end{eqnarray}
If we set the ordinate to the left hand side, $\ln\left(-\ln\left(1-W(t)\right)\right)$ and abscissa to $\ln(t)$, and apply a linear regression, we shall have a linear function that will naturally have an intercept($\mathcal{I}$) and a slope($\mathcal{S}$). Using these estimates we can generate the estimates of the shape parameter ($\hat{g}$) as well as the scale parameter ($\hat{y}$) as \st{follows} shown below,
\begin{eqnarray}
\hat{g} = \mathcal{S}, \ \ \hat{y} = \exp{\left(-\mathcal{I}/\hat{g}\right)}. \label{eqn422} 
\end{eqnarray}

In other words, the slope of the line shall be the shape parameter whereas the scale parameter needs to be calculated based on the estimate of the shape parameter according to equation \eqref{eqn422}. To demonstrate the accuracy of the Weibull assumption, we recorded around 40000 robot exchanges before they cease operation. These equal-quality robots are operating inside Quantum Scalar i6K enterprise libraries which can house up to 12000 cartridges and is optimized for high density. This data is plotted in Fig. \ref{fig:data_robot_xph} based on the formulation given in equation \eqref{eqn411}, where the intercept and slope can easily be found and used to calculate the shape parameter, $\hat{g}=0.76$ and scale parameter $\hat{y}=491669$. The accumulation in the data for $t$ values satisfying $9 \leq \ln(t) \leq 12$ is due to the fact that most robots have a logarithmic lifetime in that range. Based on the estimated parameters of the Weibull distribution, the average number of exchanges can be calculated to be $\hat{y}\Gamma(1 + 1/\hat{g}) = 580747$ exchanges before critical robot failure happens. In the numerical results section, we shall choose our parameters within the ballpark of these figures to make our results/conclusions realistic. We finally note that the use of distributions with more parameters could approximate the data better, however  it will only result in extremely minor accuracy advantage at the expense of increased estimation complexity.

\section{Numerical Results}

As it is usually the case with cold (and archival) storage platforms, we primarily focus on the read-back or in other words the data retrieval performance in this section. We present few numerical results for the proposed simulation and modeling platform. The intention is to illustrate reliability (in terms of MTTDL) and unavailability (in terms of MTTDU) in the same plot on the ordinate as a function of other simulation parameters. The abscissa  could be one of the parameters of the system including the exchange and carrier repair rates. Since the number of simulation parameters are plenty and it is hard to visualize/plot higher dimensional data using two dimensions, we present a 2-D plot where we fix most of the simulation parameters except the exchange/swap ($xph$) and carrier repair rates ($\phi$). The former typically changes based on the system utilization rate whereas the latter is under the control of system maintenance team. Another reason for choosing these parameters to vary is that they directly affect the unavailability of the system i.e., in the absence of the carrier (failed carrier) overall data access time increases until carrier repair takes over. We particularly note that most of the carrier devices (e.g. robots) are shipped with a maximum exchange/swap rate number for reliable operation (such as 840 exchanges per hour ($xph$) \cite{richards2018}), we vary the abscissa from some small exchange number to somewhere above the reported maximums and present results in a log-log plot. Similarly, when we plot the MTTDU in terms of $\phi$, we fixed the exchange/swap rate and varied the carrier repair rate to see the effect of repair frequency on the unavailability performance. Please note that the system could be operating at any point on these performance curves at a given time $t$.

\begin{table}
    \normalsize
    \caption{Simulation parameters}
    \begin{tabularx}{\columnwidth}{X|X}
        \hline
        Parameter                 & Value    \\
        \hline
        $\lambda$ (hours) & 1/50000 \\
        $\mu$ (hours)    & 1/24       \\
        $\theta$ (hours)       & 1/8760 \\
        $(n,k)$         & Variable \\
        $\epsilon$ (UCER)    & $10^{-19}$      \\
        $tape\_capacity$ & 6TB \\
        $\hat{g}$ (shape, Weibull) & 0.37 and 0.67 \\
        $\hat{y}$ (scale, Weibull) & 525985 \\
        $\kappa$ & 0.001 \\
        $\#$ of simulations & $>$10000 \\
        \hline
    \end{tabularx}
    \label{table:simulation parameters}
\end{table}

The parameters of the simulation are briefly summarized in Table \ref{table:simulation parameters}. As can be seen, we have assumed a day-long mean data/tape repair and a year-long mean failure detection time as a starting point. These numbers again are application specific and can be changed per use case. We have selected few example half-rate code parameters such as $(4,2)$ and $(6,3)$ with varying reliability guarantees. The scale--out system consists of $n$ identical libraries where each library stores and handles only one chunk of data when requested. Each library has their own unique robot and contains multiple storage units such as tapes. Tapes and robots are assumed to be of equal quality and type. The direct effect of inner details of the scale-out library system such as the total number of libraries $M$, the number of tapes per library, the geometry of the tape shelf locations etc. are accounted by the Weibull  parameter estimations (shape and scale) and data-assisted modeling framework. This data analysis saves us from getting into the inner complexities of library systems and provides us the statistical nature of the number of exchanges per library. This is later used as an input for the proposed generalized Markov model introduced in the previous section. Finally, note that UCER is assumed to be $10^{-19}$ which is way lower than $10^{-15}$, the UCER of the known disk systems. This is due to the high data durability guarantees of the next generation tape technology \cite{dholakia2008new}.

\begin{figure}[t!]
\centering
  \includegraphics[width=\linewidth]{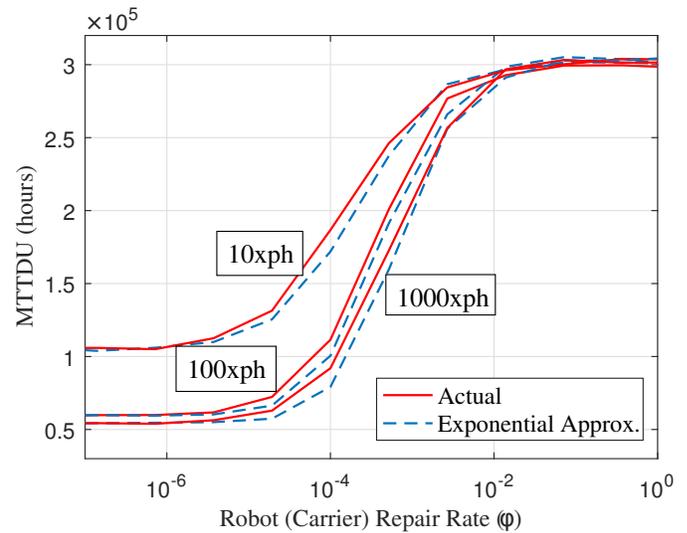}
  \caption{The accuracy of exponential-tail approximation with respect to MTTDU as a function of robot repair rate ($\phi$) using a (4,2) MDS code for two different exchange rates 10xph and 100xph.}
  \label{fig:modelling_comp}
\end{figure}

\begin{figure}[t!]
\centering
  \includegraphics[width=\linewidth]{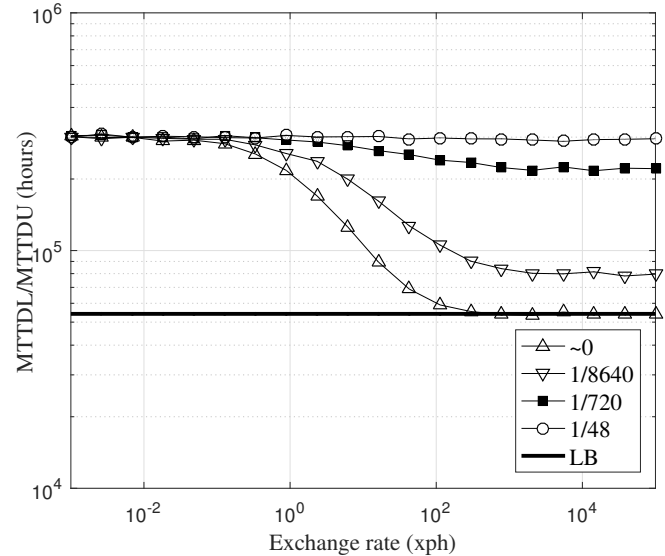}
  \caption{MTTDL/MTTDU (hours) as a function of Exchange rate (round trips or swaps per hour) and a (4,2) MDS codes with the shape parameter 0.67 for different repair rates $\phi$. LB stands for the derived lower bound.}
  \label{fig:modelling1}
\end{figure}

Our first simulation is presented in Fig. \ref{fig:modelling_comp} where we clearly demonstrate the validity of our exponential tail assumption made earlier (such as the expression \eqref{exptailapprox1}) for approximating the non-exponential distributions that appear in various stages of the proposed Markov model. We have used (4,2) MDS code and three exchange rates namely 10xph, 100xph and 1000xph and plotted MTTDU results using both the actual distributions as well as the exponential-tail approximation (used in this simulation) as a function of robot repair rate. One of the initial observations is that our exponential-tail assumption leads to a lower bound on the actual MTTDU values. Furthermore, the worst case difference between the actual and approximate MTTDU values do not affect the \textit{number of nines}, a metric typically used to express system reliability with regard to MTTDL performance metric in the industry. For the rest of this subsection, we shall present our results using the exponential-tail approximation due to simpler formulation as well as analytical tractability. 

In Fig. \ref{fig:modelling1}, we present MTTDL/MTTDU in hours as a function of exchange rate for a (4,2) MDS code. In light of our data observations and equations \eqref{eqn411} and \eqref{eqn422} derived earlier for estimating the scale and shape parameters, we have found that two shape parameters $0.37$ and $0.67$ along with the scale parameter $525985$ are most common giving us the average total number exchanges of  $695563$ and $2200634$, respectively, before a robot failure happens. Note that if a robot lasts only after one year, these numbers would indicate an average of 79.4 xph and 251.2 xph, respectively. We have also included the availability lower bound (as given by the equation \eqref{lowerbound}) in our plots which do not change with the growing exchange rate. 

As can be seen from Fig. \ref{fig:modelling1}, as the exchange rate tends to zero (going from right to left on abscissa), the reliability closely converges to the durability of the system model introduced in Fig. \ref{GMMs3} where the availability issue posses no more risk to data access anymore. On the other hand, as the exchange rate tends to large values (going from left to right on abscissa), MTTDU converges to the durability lower bound. Depending on the operating point of the library system, our model clearly shows how the unavailability changes as a function of exchange rate if we do not have sufficiently frequent robot repair in place. Also, it can be observed from the same simulation data that a robot repair rate of $\phi = 1/48$ seems to be sufficiently frequent for the system maintenance and hence we do not see any notable  reduction in the availability for this particular  repair rate.

\begin{figure}[t!]
\centering
  \includegraphics[width=\linewidth]{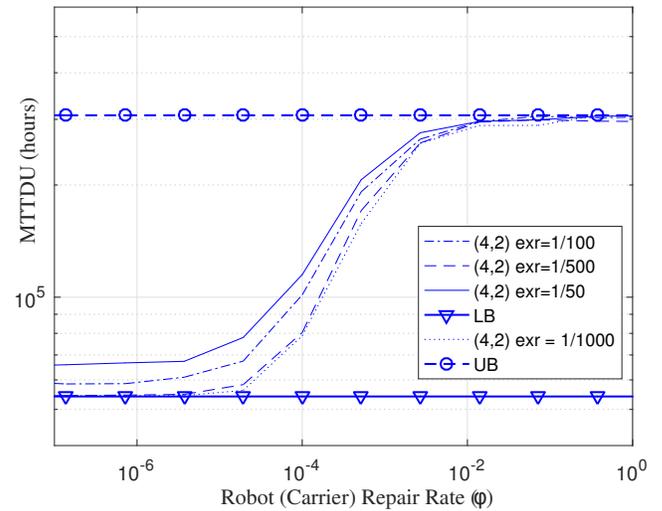}
  \caption{MTTDU (hours) as a function of robot repair rate ($\phi$) and few example exchange rates (exr) shown for (4,2) MDS code. LB stands for the lower bound.}
  \label{fig:modelling_phi1}
\end{figure}

\begin{figure}[t!]
\centering
  \includegraphics[width=\linewidth]{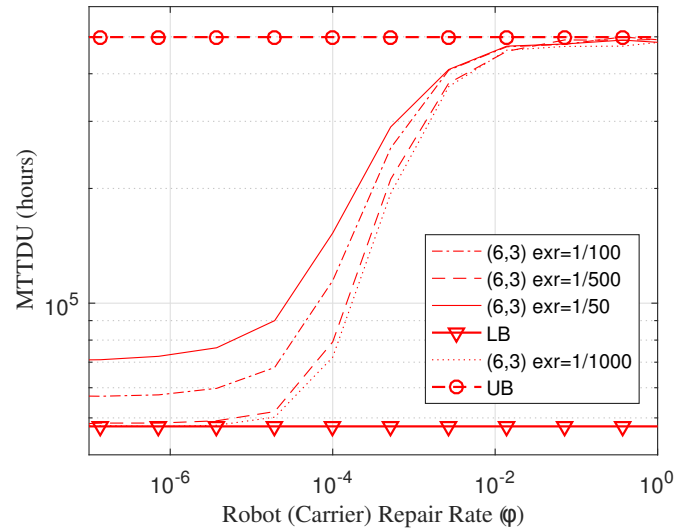}
  \caption{MTTDU (hours) as a function of robot repair rate ($\phi$) and few example exchange rates (exr) shown for (6,3) MDS code. LB stands for the lower bound.}
  \label{fig:modelling_phi2}
\end{figure}

On the other hand, we observe from Fig. \ref{fig:modelling1} (and for that matter in Figs. \ref{fig:modelling_phi1} and  \ref{fig:modelling_phi2}) that as we increase the robot repair rate $\phi$, i.e., we perform more frequent robot repairs, we improve the availability. However, at some point, increasing $\phi$ does not help us much i.e., system's robots are repaired fast enough that no unavailability leads to a dramatic performance loss.  In order to find the optimal robot repair rate, we also need to plot MTTDU as a function of $\phi$ for a range of exchange/swap rates. These performance plots are shown in Fig. \ref{fig:modelling_phi1} and Fig. \ref{fig:modelling_phi2} for the half-rate MDS codes (4,2) and (6,3), respectively. Also included in the same plots are the corresponding lower and upper bounds computed using equations \eqref{lowerbound} and \eqref{upperbound}, respectively. There are two interesting observations common to both plots. One of the observations is that MTTDU performances converge after certain exchange rates. For example in Fig. \ref{fig:modelling_phi1}, $\phi=1/500$ and $\phi=1/1000$ do not provide dramatically different MTTDU performances. Exact same trend can be observed for (6,3) MDS code in Fig. \ref{fig:modelling_phi2} as well. Therefore, since lowering the exchange rate improves the availability, we can talk about an optimal repair rate beyond which we do not experience any unavailability for all possible exchange rates of interest. Having determined the optimal exchange rate for a given system is important from both user satisfaction and energy savings point of views. The second observation with respect to these plots is that by keeping the code rate fixed, as the blocklength of the MDS code gets larger, the associated lower bound gets worse. This is due to the number of parities do not scale as much as needed to compensate for the increased blocklength. However, using the expressions derived for the lower bound (the equation \eqref{lowerbound}), one can immediately notice that the performance difference between different size MDS codes of the same rate will disappear as $n$ tends large. From Fig. \ref{fig:modelling_phi1} and Fig. \ref{fig:modelling_phi2}, we can quantify this difference for both half-rate codes, namely (4,2) and (6,3) MDS codes, respectively.

\section{Conclusion}
The reliability modeling of cold data storage involves a set of challenges due to its specific  functional requirements. Many external and data-unrelated factors play significant roles in the data-loss and data-unavailability guarantees provided to the end-user. In this study, we have pointed out several of these important factors and proposed a data-assisted general Markov model for cold data storage equipped with carrier assistance.  Furthermore, we have proposed a data-assisted simulation platform for a $(\Tilde{n},\Tilde{k})$-coded scale-out  cold data storage system in which we have accounted for different node states, hard errors and data-unavailability all at the same time. A tape library system is considered as a special use case of this reliability model. We have clearly demonstrated the effects of choosing different carrier repair rates on the reliability and availability of the system based on the operating exchange rate. In addition, useful upper and lower bounds on the availability performance are derived. Finally, we have investigated the critical choice of blocklength of the underlying fixed-rate MDS code and its effect on the system availability. One of the key features of the system is its data-driven distribution estimation framework used to model aging as well as its straightforward applicability to replication based systems for both reliability engineers and system designers. As a future work, we shall extend our model to encompass more than three node states, the possibility of having multiple carriers (robots) per node and real-time parameter estimations for an adaptive reliability model. 

\section*{Acknowledgements}
\label{Ack}
Authors would like to acknowledge various anonymous reviewers for their valuable input and suggestions that tremendously improved the quality of the presentation. This work is a joint collaboration with Quantum Corporation, LTO Advanced Development Team  and Dr. Arslan.

\appendices
\section{Upper/lower bound on $N_s$} \label{appendixA}

Observe that using Vandermonde convolution for the expression given for $N_s$, we can rewrite for it $s > 1$
\begin{align}
N_s &=  \binom{n-k+s-1}{s-1} + 1  \\
&=  \sum_{j=1}^{s-2} \binom{s-2}{j} \binom{n-k+1}{j+1} + n - k + 2 \\
&\geq \sum_{j=2}^{s-1} \left(\frac{s-2}{j-1}\right)^{j-1} \binom{n-k+1}{j} + \sum_{j=0}^{1} \binom{n-k+1}{j} \label{eqn50}
\end{align}
from which we can deduce that
\begin{eqnarray}
N_s \geq \sum_{j=0}^{s-1} \binom{n-k+1}{j}. \label{eqn51}
\end{eqnarray}
Note that the lower bound in Eq. \eqref{eqn50} is a tighter compared to one in \eqref{eqn51}. For the upper bound, we observe that
\begin{eqnarray}
N_s &=& \sum_{j=1}^{s-2} \binom{s-2}{j} \binom{n-k+1}{j+1} + n - k + 2 \nonumber \\
&=&  \sum_{j=2}^{s-1} \binom{s-2}{j-1} \binom{n-k+1}{j} + \sum_{j=0}^{1} \binom{n-k+1}{j} \label{eqn49} \nonumber \\
&\leq&  \sum_{j=0}^{s-1} \binom{s-1}{j} \binom{n-k+1}{j} = \sum_{j=0}^{s-1} \binom{s-1}{j} \binom{n-k+1}{n-k+1-j} \label{eqn50a} \nonumber \\
&=& \binom{s+n-k}{s-1}
\end{eqnarray}
where \eqref{eqn50a} results from \eqref{eqn49} using Pascal's triangle inequality which for any positive $m \geq c$ is given by
\begin{eqnarray}
\binom{m}{c} = \binom{m-1}{c-1} + \binom{m-1}{c}
\end{eqnarray}

Since $\binom{s+n-k}{s-1} = \binom{s+n-k}{n-k+1}$ we can deduce that for a fixed $s \ll n$ and a scaling  $k$ that is linear in some large $n$, i.e., $k=\alpha n$ for any $\{\alpha: 0 < \alpha < 1\}$, the total number of system (Markov) states will scale with $O(n^{\min\{s-1,n-k\}}) = O(n^{\min\{s-1,(1-\alpha)n\}})$ = $O(n^{s-1})$. In other words, the complexity of our simulation framework (and the corresponding Markov chain) grows exponentially in the number of node states unless the rate of the code goes to unity ($r \rightarrow 1$), i.e., $k$ becomes sublinear in $n$ with constant $n-k < s$. In that case the total number of system states would scale with $O(n^{n-k})$ with no dependence on the number of node states, $s$.

\section{Derivation of Performance Lower Bound} \label{appendixB}

\begin{figure*}[htp!]
\begin{tikzpicture}[node distance=1.3cm]
    %\hspace*{-1.5cm}
    \tikzset{node style/.style={state, fill=gray!20!white}}
    \tikzset{node style2/.style={state, fill=gray!80!white}}
    % Add the states
    \footnotesize
    \node[node style, ellipse] (n)   {$n$A};
    \node[node style, ellipse, right=of n] (n1) {$(n-1)$A,$1$F};
    \node[state,  line width=0.4mm, right=of n1, draw=none] (n2) {$\dots$};
    \node[node style, ellipse, right=of n2] (j) {$(k+1)$A,$(n-k-1)$F};
    \node[node style, ellipse, right=of j] (m1) {$k$A,$(n-k)$F};
    \node[node style, ellipse,  right=of m1] (m) {$\textbf{F}$};

    % Connect the states with arrows
    \draw[every loop]
        (n) edge[bend left=0, auto=left] node {$n\lambda \Delta_n$} (n1)
        (n) edge[bend right=27, auto=left] node {$n\lambda (1-
        \Delta_n)$} (m)
        (n1) edge[bend right=23, auto=left]  node {$(n-1)\lambda (1-\Delta_{n-1})$} (m)
        (n1) edge[auto=left] node {$(n-1)\lambda \Delta_{n-1}$} (n2)
        (n2) edge[auto=left] node {} (j)
        (j) edge[auto=left] node {$(k+1)\lambda \Delta_{k+1}  \atop$} (m1)
        (m1) edge[auto=left] node {} (m)
		(j) edge[bend right=18, auto=left]  node {$(k+1)\lambda (1-\Delta_{k+1})$} (m);
\end{tikzpicture}
\caption{The general CT Markov model reduces to single dimensional one with the following transition rates.}\label{fig:markov2422}
\end{figure*}
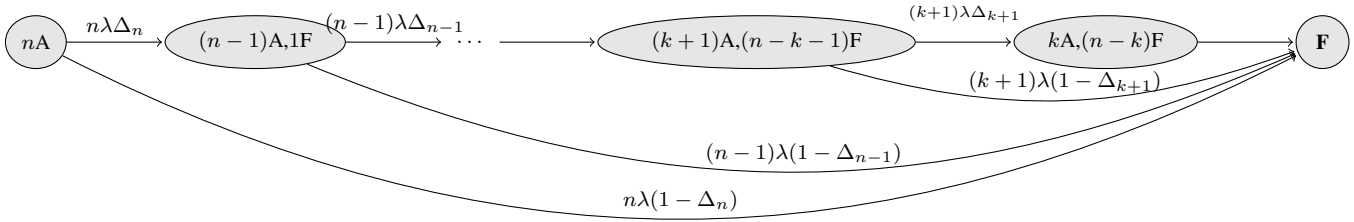

Note that we can rewrite the conditional probability that we do not end up in the total data loss (failure) state when there are $i$ available nodes, $\Delta_i$ in an explicit integral form
\begin{eqnarray}
\Delta_i &=& \frac{\Gamma(i+1)}{\Gamma(i-k)\Gamma(k+1)} \int_{\eta(t)}^1 v^{a-1} (1-v)^{b-1}dv \nonumber \\
&=& \frac{i!}{(i-k-1)!k!} \int_{\eta}^1 v^{i-k-1} (1-v)^{k}dv \nonumber \\
&=& (i-k) \binom{i}{k} \int_{\epsilon+\kappa - \epsilon\kappa}^1 v^{i-k-1} (1-v)^{k}dv
\end{eqnarray}
where $\eta = 1 - (1-\epsilon)(1-\kappa)$ is the hard error rate. To derive the lower bound we consider the case of total carrier unavailability. Thus, this results in no failure detection and henceforth no data repair process is initiated. This leads to the simplified version of the Markov model as shown in Fig. \ref{fig:markov2422}. Note that the hard error that leads to total failure when there are $i$ available nodes happens with probability $(1-\Delta_i)\prod_{j=i+1}^n \Delta_j$. The total time before failure is the sum of average times spent in each of the visited system states i.e., $1/n\lambda, 1/(n-1)\lambda, \dots, 1/i\lambda$. Finally, by summing over all possible $i$, we can estimate the lower bound on the total average time spent before failure as
\begin{eqnarray}
LB &=& \sum_{i=k}^n (1 - \Delta_i) \prod_{j=i+1}^n \Delta_j \sum_{j=i}^n \frac{1}{j\lambda}  \nonumber \\
&=& \sum_{i=k}^n (1 - \Delta_i) \prod_{j=i+1}^n \Delta_j \left( hs(n) - hs(i-1) \right) \nonumber \\
&\approx& \sum_{i=k}^n (1 - \Delta_i) \prod_{j=i+1}^n \Delta_j  \Bigg( \log\left(\frac{n}{i-1}\right)^{1/\lambda}   \nonumber \\ && + \frac{1}{\lambda}\sum_{l=0}^1(-1)^l\left(\frac{1}{2n_l} - \frac{1}{12n_l^2} + \frac{1}{120n_l^4}\right) \Bigg) 
\end{eqnarray}
with $\Delta_k = 0$ and $n_l = n - l(n-i+1)$. Note  that if there is no hard errors, i.e., $\Delta_i = 1$ for $i=n,\dots,k+1$, we shall have the simplified lower bound with the closed form expression given by $\sum_{j=k}^n 1/j \lambda$. In general we have
\begin{eqnarray}
LB \leq \sum_{j=k}^n \frac{1}{j \lambda} =hs(n) - hs(k-1)
\end{eqnarray}
Note that the approximation follows due to  \eqref{eq_approx}.

\ifCLASSOPTIONcaptionsoff
  \newpage
\fi



%\bibliographystyle{IEEEtran}

%\section*{References}

\bibliography{main.bib}{}

\begin{thebibliography}{10}
\providecommand{\url}[1]{#1}
\csname url@samestyle\endcsname
\providecommand{\newblock}{\relax}
\providecommand{\bibinfo}[2]{#2}
\providecommand{\BIBentrySTDinterwordspacing}{\spaceskip=0pt\relax}
\providecommand{\BIBentryALTinterwordstretchfactor}{4}
\providecommand{\BIBentryALTinterwordspacing}{\spaceskip=\fontdimen2\font plus
\BIBentryALTinterwordstretchfactor\fontdimen3\font minus
  \fontdimen4\font\relax}
\providecommand{\BIBforeignlanguage}[2]{{%
\expandafter\ifx\csname l@#1\endcsname\relax
\typeout{** WARNING: IEEEtran.bst: No hyphenation pattern has been}%
\typeout{** loaded for the language `#1'. Using the pattern for}%
\typeout{** the default language instead.}%
\else
\language=\csname l@#1\endcsname
\fi
#2}}
\providecommand{\BIBdecl}{\relax}
\BIBdecl

\bibitem{allen2017materiality}
J.~Allen-Robertson, ``{The Materiality of Digital Media: The Hard Disk Drive,
  Phonograph, Magnetic Tape and Optical Media in Technical Close-up},''
  \emph{New Media \& Society}, vol.~19, no.~3, pp. 455--470, 2017.

\bibitem{church2012next}
G.~M. Church, Y.~Gao, and S.~Kosuri, ``{Next-Generation Digital Information
  Storage in DNA},'' \emph{Science}, vol. 337, no. 6102, pp. 1628--1628, 2012.

\bibitem{Qing2018}
Y.~Qing, S.~A. Ionescu, G.~S. Pulcu, and H.~Bayley, ``{Directional Control of a
  Processive Molecular Hopper},'' \emph{Science}, vol. 361, no. 6405, pp.
  908--912, 2018.

\bibitem{arslan2014mds}
S.~S. Arslan, J.~Lee, J.~Hodges, J.~Peng, H.~Le, and T.~Goker, ``{MDS Product
  Code Performance Estimations under Header CRC Check Failures and Missing
  Syncs},'' \emph{IEEE Transactions on Device and Materials Reliability,},
  vol.~14, no.~3, pp. 921--930, 2014.

\bibitem{oprea2010clean}
A.~Oprea and A.~Juels, ``A {C}lean-{S}late {L}ook at {D}isk {S}crubbing.'' in
  \emph{FAST}, 2010, pp. 57--70.

\bibitem{ryu2009effects}
J.~Ryu and C.~Park, ``Effects of {D}ata {S}crubbing on {R}eliability in
  {S}torage {S}ystems,'' \emph{IEICE Transactions on Information and Systems},
  vol.~92, no.~9, pp. 1639--1649, 2009.

\bibitem{schwarz2004disk}
T.~J. Schwarz, Q.~Xin, E.~L. Miller, D.~D. Long, A.~Hospodor, and S.~Ng, ``Disk
  {S}crubbing in {L}arge {A}rchival {S}torage {S}ystems,'' in \emph{The IEEE
  Computer Society's 12th Annual International Symposium on Modeling, Analysis,
  and Simulation of Computer and Telecommunications Systems, 2004.(MASCOTS
  2004). Proceedings.}\hskip 1em plus 0.5em minus 0.4em\relax IEEE, 2004, pp.
  409--418.

\bibitem{iliadis2008disk}
I.~Iliadis, R.~Haas, X.-Y. Hu, and E.~Eleftheriou, ``Disk {S}crubbing versus
  {I}ntra-{D}isk {R}edundancy for {H}igh-{R}eliability {RAID} {S}torage
  {S}ystems,'' in \emph{ACM SIGMETRICS Performance Evaluation Review}, vol.~36,
  no.~1.\hskip 1em plus 0.5em minus 0.4em\relax ACM, 2008, pp. 241--252.

\bibitem{Hafner2006}
J.~L. Hafner and K.~Rao, ``Notes on {R}eliability {M}odels for {N}on-{MDS}
  {E}rasure {C}odes,'' \emph{IBM Res. rep. RJ--10391, 2006}, 2006.

\bibitem{dholakia2008new}
A.~Dholakia, E.~Eleftheriou, X.-Y. Hu, I.~Iliadis, J.~Menon, and K.~Rao, ``A
  {N}ew {I}ntra-{D}isk {R}edundancy {S}cheme for {H}igh-{R}eliability {RAID}
  {S}torage {S}ystems in the {P}resence of {U}nrecoverable {E}rrors,''
  \emph{ACM Transactions on Storage (TOS)}, vol.~4, no.~1, p.~1, 2008.

\bibitem{Paris2012}
J.-F. P{\^a}ris, S.~T.~J. Schwarz, A.~Amer, and D.~D. Long, ``Highly {R}eliable
  {T}wo-{D}imensional {RAID} {A}rrays for {A}rchival {S}torage,'' in \emph{2012
  IEEE 31st International Performance Computing and Communications Conference
  (IPCCC)}.\hskip 1em plus 0.5em minus 0.4em\relax IEEE, 2012, pp. 324--331.

\bibitem{Arslan2018}
S.~S. Arslan, ``{A Reliability Model for Dependent and Distributed MDS Disk
  Array Units},'' \emph{IEEE Transactions on Reliability}, vol.~68, no.~1, pp.
  133--148, 2019.

\bibitem{arslan2014redundancy}
------, ``{Redundancy and Aging of Efficient Multidimensional MDS
  Parity-Protected Distributed Storage Systems},'' \emph{IEEE Transactions on
  Device and Materials Reliability}, vol.~14, no.~1, pp. 275--285, 2014.

\bibitem{rahman2018analysis}
P.~Rahman and G.~D. N.~F. Shavier, ``{Analysis of Mean Time to Data Loss of
  Fault-Tolerant Disk Arrays RAID-6 based on Specialized Markov Chain},'' in
  \emph{IOP Conference Series: Materials Science and Engineering}, vol. 327,
  no.~2.\hskip 1em plus 0.5em minus 0.4em\relax IOP Publishing, 2018, pp.
  022--086.

\bibitem{hall2016tools}
R.~J. Hall, ``Tools for predicting the reliability of large-scale storage
  systems,'' \emph{ACM Transactions on Storage (TOS)}, vol.~12, no.~4, p.~24,
  2016.

\bibitem{constantopoulos2005}
P.~Constantopoulos, M.~Doerr, and M.~Petraki, ``Reliability {M}odeling for
  {L}ong {T}erm {D}igital {P}reservation,'' in \emph{9th DELOS Network of
  Excellence Thematic Workshop “Digital Repositories: Interoperability and
  Common Services”, Foundation for Research and Technology-Hellas (FORTH)},
  2005.

\bibitem{HanChan}
Y.~Han and C.~P. Chan, ``{Modeling System Reliability for Digital Preservation:
  Model Modification and Four-Copy Model Study},'' in \emph{Proceedings of the
  Fifth International Conference on Preservation of Digital Objects
  (iPRES)}.\hskip 1em plus 0.5em minus 0.4em\relax The British Library,
  London., 2008.

\bibitem{ivanichkina2015}
L.~Ivanichkina and A.~Neporada, ``{The Reliability Model of a Distributed Data
  Storage in Case of Explicit and Latent Disk Faults},'' \emph{ARPN Journal of
  Engineering and Applied Sciences.--10 (20).--2015.--C}, pp. 9150--9158, 2015.

\bibitem{Judge2003}
J.~Judge, R.~Schmidt, R.~Weiss, and G.~Miller, ``{Media Stability and Life
  Expectancies of Magnetic Tape for Use with IBM 3590 and Digital Linear Tape
  systems},'' in \emph{20th IEEE/11th NASA Goddard Conference on Mass Storage
  Systems and Technologies, 2003.(MSST 2003). Proceedings.}\hskip 1em plus
  0.5em minus 0.4em\relax IEEE, 2003, pp. 97--100.

\bibitem{weiss2002}
R.~D. Weiss, ``{Environmental Stability Study and Life Expectancies of Magnetic
  Media for Use with IBM 3590 and Quantum Digital Linear Tape Systems.}''
  \emph{Report to National Archives and Records Administration
  \#NAMA-01-F-0061}, 2002.

\bibitem{shahani2003}
C.~J. Shahani, B.~Manns, and M.~Youket, ``{Longevity of CD media: research at
  the library of congress},'' \emph{Preservation Research and Testing Division,
  Washington, DC}, 2003.

\bibitem{jernigan2009data}
R.~P. Jernigan~IV, A.~Tracht, and P.~F. Corbett, ``{Data Allocation within A
  Storage System Architecture},'' Nov.~10 2009, {U}S Patent 7,617,370.

\bibitem{cidon2013copysets}
A.~Cidon, S.~Rumble, R.~Stutsman, S.~Katti, J.~Ousterhout, and M.~Rosenblum,
  ``{Copysets: Reducing the frequency of data loss in cloud storage},'' in
  \emph{2013 USENIX Annual Technical Conference}, 2013, pp. 37--48.

\bibitem{botezatu2016predicting}
M.~M. Botezatu, I.~Giurgiu, J.~Bogojeska, and D.~Wiesmann, ``{Predicting Disk
  Replacement towards Reliable Data Centers},'' in \emph{Proceedings of the
  22nd ACM SIGKDD International Conference on Knowledge Discovery and Data
  Mining}.\hskip 1em plus 0.5em minus 0.4em\relax ACM, 2016, pp. 39--48.

\bibitem{mahdisoltani2017proactive}
F.~Mahdisoltani, I.~Stefanovici, and B.~Schroeder, ``{Proactive error
  prediction to improve storage system reliability},'' in \emph{2017 USENIX
  Annual Technical Conference}, 2017, pp. 391--402.

\bibitem{burkhard1993}
W.~A. Burkhard and J.~Menon, ``{Disk array storage system reliability},'' in
  \emph{FTCS-23 The Twenty-Third International Symposium on Fault-Tolerant
  Computing}.\hskip 1em plus 0.5em minus 0.4em\relax IEEE, 1993, pp. 432--441.

\bibitem{wideman2016data}
R.~B. Wideman, S.~S. Arslan, J.~Lee, and T.~Goker, ``{Data Deduplication with
  Adaptive Erasure Code Redundancy},'' Nov.~22 2016, {US Patent 9,503,127}.

\bibitem{suayb2019joint}
S.~S. Arslan, T.~Goker, and R.~B. Wideman, ``{Joint De-Duplication-Erasure
  Coded Distributed Storage},'' Jun. 11 2019, {US} Patent 10,318,389.

\bibitem{LRC}
D.~S. Papailiopoulos and A.~G. Dimakis, ``{Locally Repairable Codes},''
  \emph{IEEE Transactions on Information Theory}, vol.~60, no.~10, pp.
  5843--5855, 2014.

\bibitem{dimakis2011survey}
A.~G. Dimakis, K.~Ramchandran, Y.~Wu, and C.~Suh, ``{A Survey on Network Codes
  for Distributed Storage},'' \emph{Proceedings of the IEEE}, vol.~99, no.~3,
  pp. 476--489, 2011.

\bibitem{Satyanarayanan}
M.~Satyanarayanan, ``{A Study of File Sizes and Functional Lifetimes},'' in
  \emph{ACM SIGOPS Operating Systems Review}, vol.~15, no.~5.\hskip 1em plus
  0.5em minus 0.4em\relax ACM, 1981, pp. 96--108.

\bibitem{downey2001structural}
A.~B. Downey, ``{The Structural Cause of File Size Distributions},'' in
  \emph{MASCOTS 2001, Proceedings Ninth International Symposium on Modeling,
  Analysis and Simulation of Computer and Telecommunication Systems}.\hskip 1em
  plus 0.5em minus 0.4em\relax IEEE, 2001, pp. 361--370.

\bibitem{ramaswami2014modeling}
V.~Ramaswami, K.~Jain, R.~Jana, and V.~Aggarwal, ``{Modeling Heavy Tails in
  Traffic Sources for Network Performance Evaluation},'' in \emph{Computational
  Intelligence, Cyber Security and Computational Models}.\hskip 1em plus 0.5em
  minus 0.4em\relax Springer, 2014, pp. 23--44.

\bibitem{gong2001tails}
W.~Gong, Y.~Liu, V.~Misra, and D.~Towsley, ``{On the Tails of Web File Size
  Distributions},'' in \emph{Proceedings of the Annual Allerton Conference on
  Communication Control and Computing}, vol.~39, no.~1.\hskip 1em plus 0.5em
  minus 0.4em\relax The University; 1998, 2001, pp. 192--201.

\bibitem{richards2018}
S.~Richards, ``{Maintaining a Large Scale, Very Active Tape Archive},'' in
  \emph{34th International Conference on Massive Storage Systems and
  Technology}.\hskip 1em plus 0.5em minus 0.4em\relax IEEE, 2018.

\end{thebibliography}
\bibliographystyle{IEEEtran}

%\bibliographystyle{ieeetran}
%\bibliography{mybibfile}
%\bibliography{/bib/IEEEexample}

%\begin{thebibliography}{1}

%H.~Kopka and P.~W. Daly, \emph{A Guide to \LaTeX}, 3rd~ed.\hskip 1em plus 0.5em minus 0.4em\relax Harlow, England: Addison-Wesley, 1999.

%\end{thebibliography}

%\end{IEEEbiography}


%\newpage



%\vfill


\end{document}